\documentclass{IET}

\usepackage{hyperref} 
\usepackage{graphics} 
\usepackage[english]{babel}
\usepackage[utf8]{inputenc}
\usepackage{algorithm}

\usepackage[]{algpseudocode}

\algnewcommand\algorithmicinput{\textbf{Input:}}
\algnewcommand\Input{\item[\algorithmicinput]}
\algnewcommand\algorithmicoutput{\textbf{Output:}}
\algnewcommand\Output{\item[\algorithmicoutput]}
\algnewcommand\algorithmichline{}
\algnewcommand\Hline{\item[\algorithmichline]}
\makeatletter
    \newcommand*{\algrule}[1][\algorithmicindent]{\makebox[#1][l]{\hspace*{.5em}\thealgruleextra\vrule height \thealgruleheight depth \thealgruledepth}}%
\newcommand*{\thealgruleextra}{}
\newcommand*{\thealgruleheight}{.85\baselineskip}
\newcommand*{\thealgruledepth}{.25\baselineskip}

\newcount\ALG@printindent@tempcnta
\def\ALG@printindent{%
    \ifnum \theALG@nested>0
        \ifx\ALG@text\ALG@x@notext
        \else
            \unskip
            \addvspace{-1pt}
            \ALG@printindent@tempcnta=1
            \loop
                \algrule[\csname ALG@ind@\the\ALG@printindent@tempcnta\endcsname]%
                \advance \ALG@printindent@tempcnta 1
            \ifnum \ALG@printindent@tempcnta<\numexpr\theALG@nested+1\relax
            \repeat
        \fi
    \fi
    }%
\usepackage{etoolbox}
\patchcmd{\ALG@doentity}{\noindent\hskip\ALG@tlm}{\ALG@printindent}{}{\errmessage{failed to patch}}
\makeatother

\newbox\statebox
\newcommand{\myState}[1]{%
    \setbox\statebox=\vbox{#1}%
    \edef\thealgruleheight{\dimexpr \the\ht\statebox+1pt\relax}%
    \edef\thealgruledepth{\dimexpr \the\dp\statebox+1pt\relax}%
    \ifdim\thealgruleheight<.75\baselineskip
        \def\thealgruleheight{\dimexpr .75\baselineskip+1pt\relax}%
    \fi
    \ifdim\thealgruledepth<.25\baselineskip
        \def\thealgruledepth{\dimexpr .25\baselineskip+1pt\relax}%
    \fi
    \State #1%
    \def\thealgruleheight{\dimexpr .75\baselineskip+1pt\relax}%
    \def\thealgruledepth{\dimexpr .25\baselineskip+1pt\relax}%
}

\begin{document}

\title{Spectrum-Convertible BVWXC Placement in OFDM-based Elastic Optical Networks}

\author[1,*]{Mohammad Hadi}
\affil{Sharif University of Technology, Azadi St., Tehran, Iran}

\author[1]{Mohammad Reza Pakravan}

\affil[*]{mhadi@ee.sharif.edu}

\abstract{Spectrum conversion can improve the performance of OFDM-based Elastic Optical Networks (EONs) by relaxing the continuity constraint and consequently reducing connection request blocking probability during Routing and Spectrum Assignment (RSA) process. We propose three different architectures for including spectrum conversion capability in Bandwidth-Variable Wavelength Cross-Connects (BVWXCs). To compare the capability of the introduced architectures, we develop an analytical method for computing average connection request blocking probability in a spectrum-convertible OFDM-based EON in which all, part or none of the BVWXCs can convert the spectrum. An algorithm for distributing a limited number of Spectrum-Convertible Bandwidth-Variable Wavelength Cross-Connects (SCBVWXCs) in an OFDM-based EON is also proposed. Finally, we use simulation results to evaluate the accuracy of the proposed method for calculating connection request blocking probability and the capability of the introduced algorithm for SCBVWXC placement. }


\maketitle

\section{Introduction}\label{sec_I}
Coherent Optical Orthogonal Frequency Division Multiplexing (CO-OFDM) is a well-known, attractive and promising solution for implementing Elastic Optical Networks (EONs) \cite{Kazuro_Survey}. OFDM-based EONs can provide scalability, flexibility, efficiency and fine granularity in resource provisioning compared to the conventional Wavelength Division Multiplexing (WDM) networks \cite{Cugini_Survey, Bijoy_Survey,Mukherjee_Survey, Armstrong_Survey}. Enabling technologies such as Bandwidth-Variable Transponders (BVTs), Sliceable Bandwidth-Variable Transponder (SBVTs) and Bandwidth-Variable Wavelength Cross-Connects (BVWXCs) have been designed and demonstrated in experimental flexible optical network test-benches \cite{exp1,exp2} and employed in architectures such as Spectrum-sLICed Elastic optical path network (SLICE) \cite{Bijoy_Survey,SLICE1,SLICE2}.

In OFDM-based EONs, an all-optical trail with multiple consecutive spectrum slots connecting a pair of source and destination nodes is named a spectrum lightpath. Routing and Spectrum Assignment (RSA) is the main resource allocation procedure in OFDM-based EONs in which a spectrum lightpath with sufficient number of consecutive spectrum slots is allocated to each traffic demand. Spectrum continuity, spectrum contiguity, guard allocation and conflict-free spectrum assignment are the main constraints in a normal RSA problem. A blocked connection request is a connection request that cannot be accommodated in the network by the online RSA solver algorithm. Thus, connection request blocking probability arises as the main performance criterion in an online RSA algorithm \cite{Talebi_Survey,Bijoy_Survey,Mukherjee_Survey, RSA1, RSA2}. 

Spectrum conversion, which is the capability of spectrum shifting in the frequency domain, can relax the RSA spectrum continuity constraint to improve the connection request blocking probability. Spectrum conversion can be realized by all optical or Optical/Electrical/Optical (O/E/O) techniques as described in \cite{Bijoy_Survey,SC1, SC2, SC3}. Spectrum conversion capability is a fundamental feature of the advanced optical cross-connect architectures such as Architecture on Demand (AoD) \cite{aod}. Although spectrum conversion capability increases the complexity of the cross-connect architecture and introduces an additional cost, its potential ability for significantly improving the network performance may persuade us to distribute a limited number of Spectrum-Convertible BVWXCs (SCBVWXCs) in a given EON topology. We assume that all, part or none of the BVWXCs in a given EON may be equipped with spectrum conversion capability. The embedded spectrum conversion capability in a SCBVWXC can be shared among different elements of the cross-connect such as output ports \cite{WC1}. Considering the order of the spectrum conversion sharing, various architectures named Full, Share-per-Node and Share-per-Link are introduced which are all inherited from the proposed WDM-based wavelength-convertible optical switch architectures in \cite{WC1}. To numerically evaluate  the amount of the performance caused by different SCBVWXC architectures, we propose a framework for calculating the average connection request blocking probability in a given OFDM-based EON. The framework covers different scenarios distinguished by various methods of sharing and distributing spectrum conversion capability in BVWXCs and network topology, respectively. As another contribution, we propose an algorithm for distributing a certain number of SCBVWXCs in a given network topology. Finally, we use simulation results to confirm the expected performance improvement by SCBVWXCs and to evaluate the ability of the proposed algorithm for distributing SCBVWXCs in a network.

The rest of the paper is organized as follows. The methods of sharing and distributing spectrum conversion capability in a BVWXC and a given network topology are discussed in Section \ref{sec_II}. In Section \ref{sec_III}, we develop our framework for computing connection request blocking probability in an arbitrary EON topology. The algorithm for distributing SCBVWXCs in an EON is proposed in Section \ref{sec_IV}. Simulation results and conclusion are included in Sections \ref{sec_V} and \ref{sec_VI}, respectively. 

\section{Spectrum Conversion in OFDM-based EONs}\label{sec_II}
Spectrum conversion is the ability of shifting the content of a contiguous spectrum interval in the frequency domain. The proposed techniques for constructing wavelength converters in WDM networks can be used to realize SCBVWXCs \cite{SC1, SC2, SC3,WC1}. In O/E/O techniques, the optical signal is first translated into the electronic domain using a BVT or SBVT. The generated electronic signal is then used to drive another BVT or SBVT tuned to put the spectrum in the desired frequency location. The process of O/E/O conversion is complex, power hungry, expensive and may adversely affect transparency by distorting the information of phase, frequency, and analog amplitude of the optical signal during the conversion process \cite{WC1,WC2}. In all-optical spectrum conversion,  the optical signal is allowed to remain in the optical domain throughout the conversion process but it suffers from wave distortion and limited range of spectrum shifting in the frequency domain. Nonlinear effects such as Wave Mixing and Cross-Modulation can be used for all-optical implementation of SCBVWXCs\cite{WC1,WC2}. 
 
Considering the complexity and cost of spectrum conversion, different architectures and distribution methods may be proposed to share and distribute a limited amount of spectrum conversion capability in BVWXCs and a given network topology, respectively. Different architectures and distribution methods are introduced in the following two sub-sections.
\subsection{Design of SCBVWXCs}\label{sec_II_A}
\begin{figure}[t!]
\center{\includegraphics[scale=0.85]{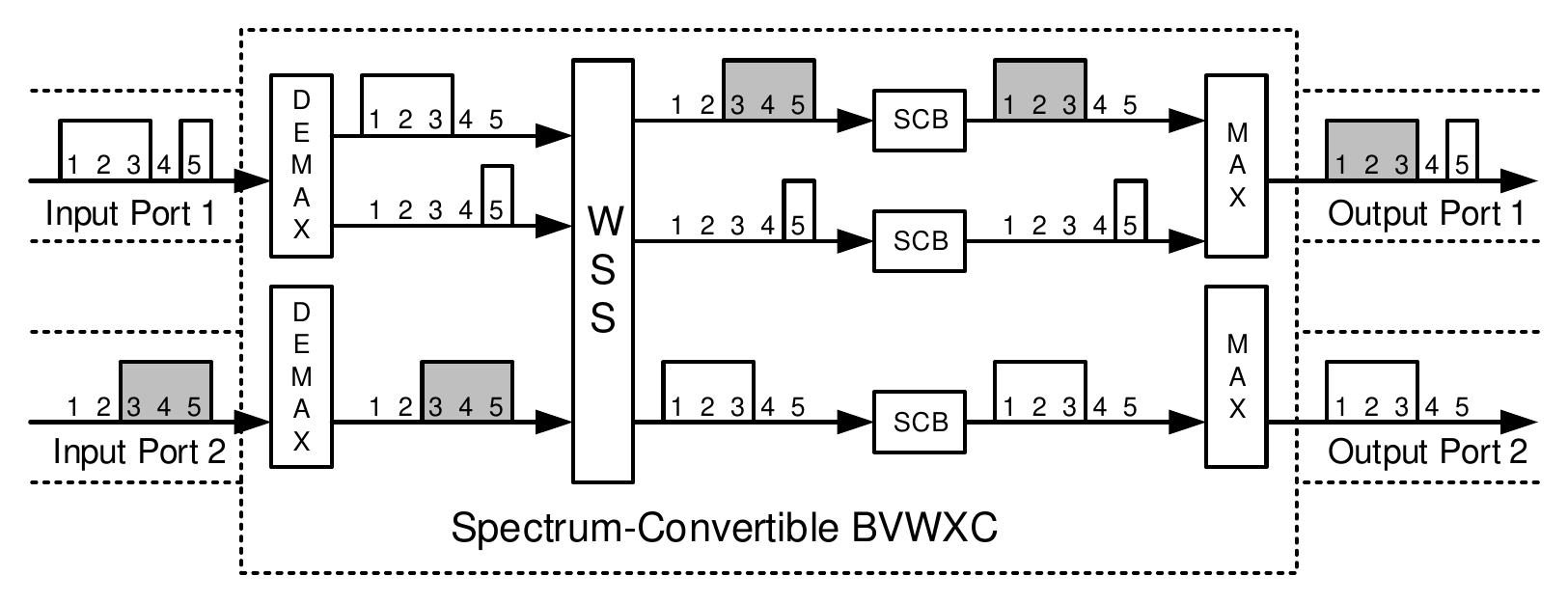}}
\center{\caption{\label{Full} A spectrum-convertible bandwidth-variable wavelength cross-connect having Full architecture.}}
\end{figure}

\begin{figure}[t!]
\center{\includegraphics[scale=0.85]{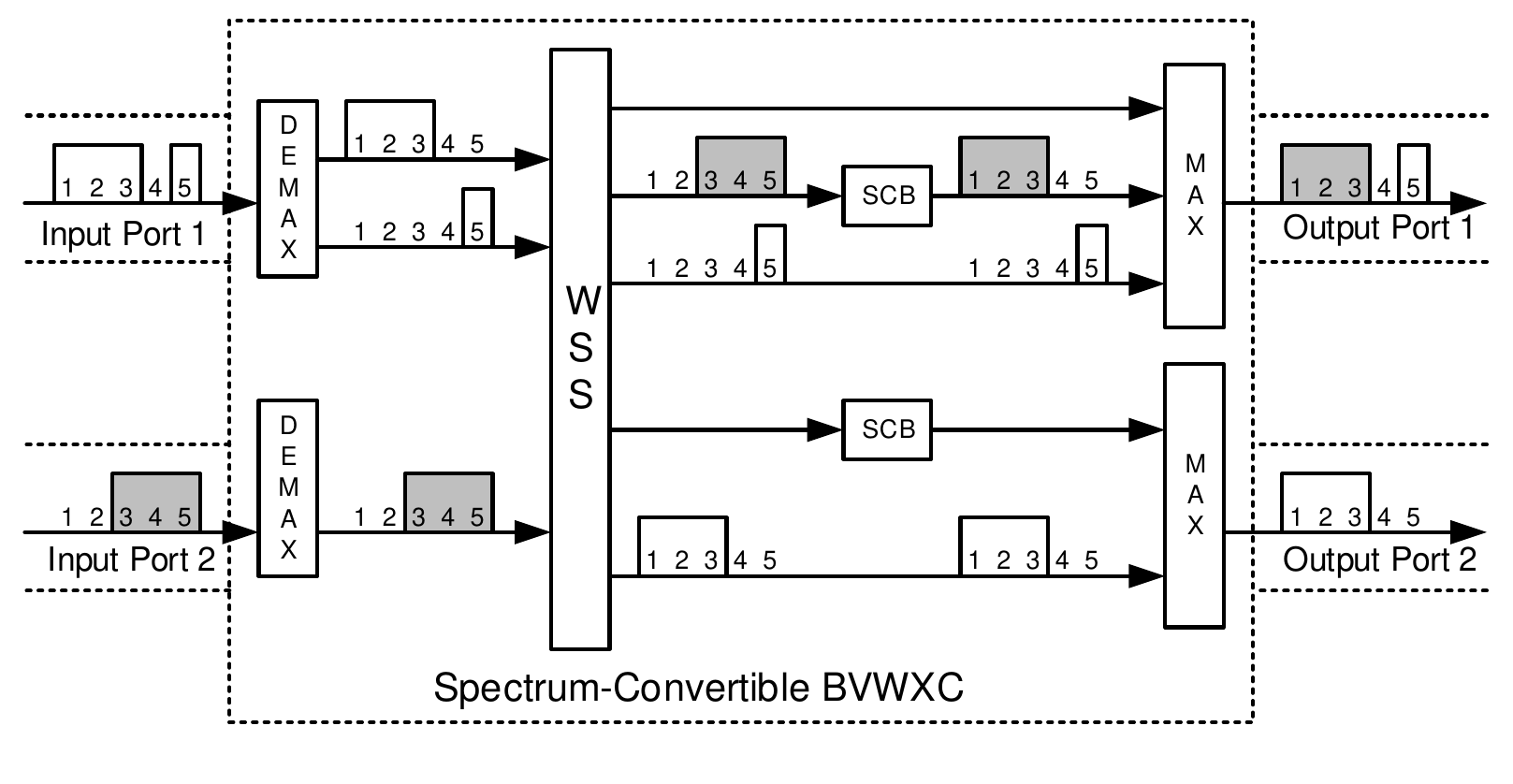}}
\center{\caption{\label{Link} A spectrum-convertible bandwidth-variable wavelength cross-connect having Share-per-Link architecture.}}
\end{figure}

\begin{figure}[t!]
\center{\includegraphics[scale=0.85]{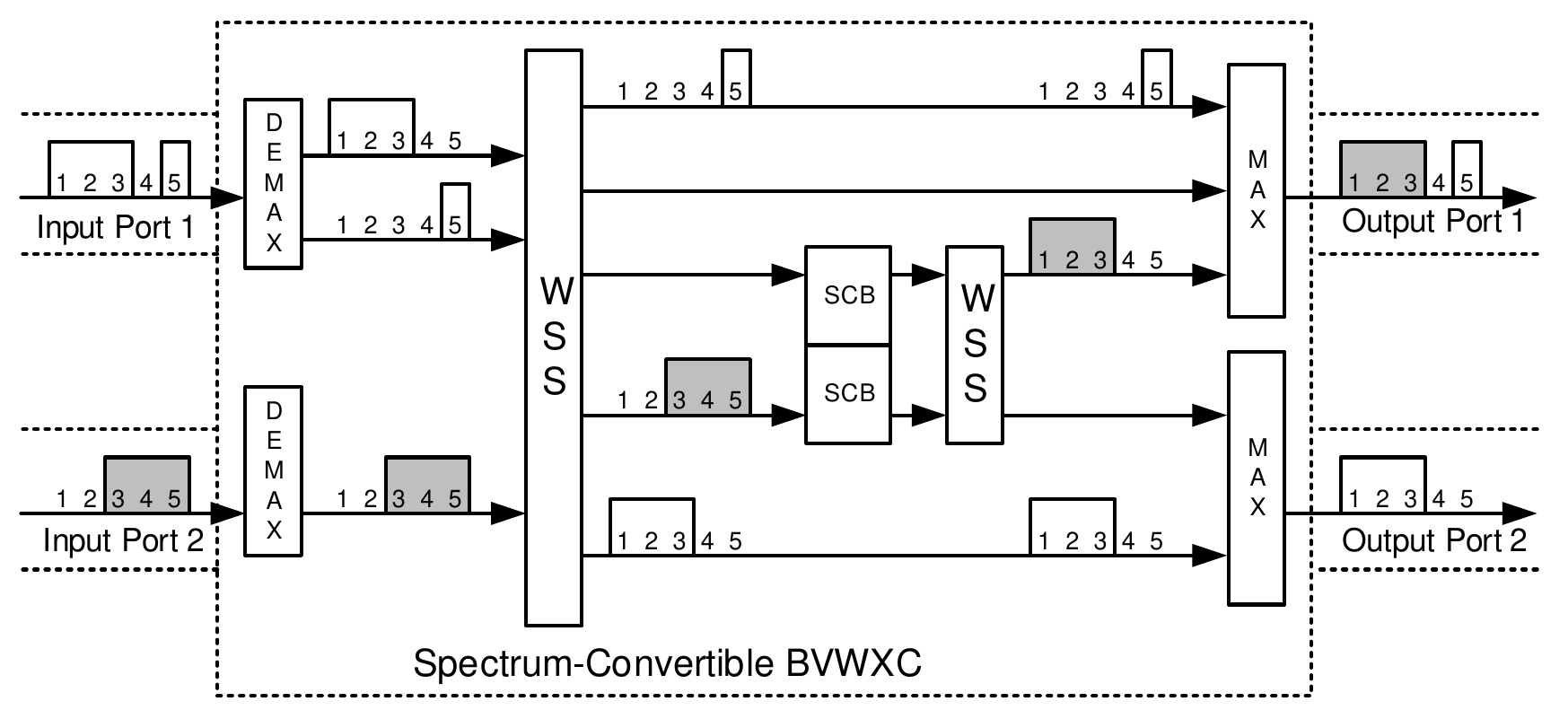}}
\center{\caption{\label{Node} A spectrum-convertible bandwidth-variable wavelength cross-connect having Share-per-Node architecture.}}
\end{figure}

We define a Spectrum Converter Box (SCB) as a module that accepts an incoming spectrum of a lightpath and shifts it to a desired spectrum band. We assume SCB is an ideal module that can convert the spectrum without any impairments such as distortion and attenuation. According to how SCBs are shared in a BVWXC, we adopt the proposed architectures in \cite{WC1} to introduce three SCBVWXC architectures named Full, Share-per-Link and Share-per-Node. In Full architecture, all lightpaths crossing the SCBVWXC have their individual SCBs. Although, Spectrum continuity constraint is fully relaxed between incoming and outcoming links of a SCBVWXC having Full architecture, the high cost prevents commercial realization of such architecture. To break the cost, a bank of SCBs can be shared among all the lightpaths crossing the SCBVWXC. We refer to this architecture as Share-per-Node architecture. Compared to Full architecture, Share-per-Node architecture provides lower spectrum conversion capability for decreased cost of architecture. In Share-per-Link architecture, each output link has its own bank of SCBs. Obviously, the spectrum conversion capability and cost of Share-per-Link architecture falls between two other architectures. Fig. \ref{Full}, Fig. \ref{Link} and Fig. \ref{Node} show the block diagram of the three introduced architectures.
\subsection{Distribution of SCBVWXCs}\label{sec_II_B}
Assume a given network topology without spectrum conversion capability. We say this network has Empty distribution of SCBVWXCs. The spectrum continuity constraint must completely be held in a network topology with Empty distribution. On the other hand, in Full distribution, all of the BVWXCs in the network topology have Full architecture. In such situation, the spectrum continuity constraint is totally relaxed in the network. In a more practical and general condition named Sparse distribution, part of the BVWXCs may be equipped with the spectrum conversion capability. All of the SCBVWXCs in Sparse distribution may have same architecture or each of them may arbitrarily have one of the mentioned architectures. Undoubtedly, Empty and Full distributions are special cases of Sparse distribution. Tab. \ref{Fig_Scenarios} shows the various scenarios considered in this paper which are distinguished by the introduced architectures and distributions of SCBVWXCs.
\section{Computational Framework for Connection Request Blocking Probability}\label{sec_III}
We define a network as a graph $G(V, E)$ where $V$ represents the set of optical nodes and $E$ represents the set of directional fiber links. Each fiber has $F$ spectrum slots. A connection request from source $s$ to destination $d$ is specified by notation $C_{sd} = (R_{sd}, T_{sd}, S_{sd})$ where the first element refers to request rate which is assumed to be Poisson with mean $R_{sd}$, the second element refers to connection hold time which is assumed to be exponential with mean $T_{sd}$ and the last element stands for number of required spectrum slots which is assumed to be an arbitrary distribution with probability mass function $P_{sd}(S)$ and mean $S_{sd}$. Characterizing parameters of a connection request are assumed to be independent of each other and other connection requests. Obviously, for each connection request, the number of required spectrum slots has Compound Poisson distribution with mean $R_{sd}T_{sd}S_{sd}$. Any connection request is routed using Shortest Path algorithm and Random Fit algorithm is used for spectrum allocation which randomly assigns one of the possible contiguous spectrum packages to the connection request \cite{Bijoy_Survey,Mukherjee_Survey,RSA1,RSA2}. 

Referring to \cite{SC1}, we follow a two stage procedure to develop our framework for calculating connection request blocking probability. Firstly, we develop a computational framework for computing the blocking probability of an $H$-hop lightpath request and then extend it to work for a given network topology.

\subsection{Computational Framework for An $H$-hop Lightpath}\label{sec_III_A}
In this section, a computational framework for calculating the blocking probability of an $H$-hop end-to-end lightpath request is developed. To make our framework, we begin with the first scenario in Tab. \ref{Fig_Scenarios} and proceed step by step to get the most general case i.e. sixth scenario. Note that various scenarios in Tab. \ref{Fig_Scenarios} are distinguished by unique ID numbers.
\begin{table}
\center
\caption{\label{Fig_Scenarios} Different scenarios considered in the paper. Each row characterizes a scenario by declaring its distribution, architecture and assigned unique ID number.}
\begin{tabular}{|c|c|c|}
\hline
\textbf{Distribution} & \textbf{Architecture} & \textbf{ID}\\
\hline
Empty & No conversion capability for all BVWXCs & $1$ \\
\hline 
Full & Full architecture for all BVWXCs & $2$ \\
\hline 
Sparse & Full architecture for all SCBVWXCs & $3$ \\
\hline 
Sparse & Share-per-Link architecture for all SCBVWXCs & $4$ \\
\hline 
Sparse & Share-per-Node architecture for all SCBVWXCs & $5$ \\
\hline 
Sparse & Arbitrary architecture for each BVWXC & $6$ \\
\hline

\end{tabular}
\end{table}

\subsubsection{$H$-hop Lightpath Connection Request Blocking Probability in First Scenario}\label{sec_III_A_1}
Assume $P_{B}^{sd}(S)$ shows blocking probability of a connection request that requires $S$ spectrum slots, originates from node $s$, travels on an $H$-hop shortest path and terminates in node $d$. The hops and nodes of the $H$-hop shortest path are numbered according to the convention shown in Fig. \ref{Numbering_SC}. We assume spectrum slots of $h$th link of the path are independently free with probability $\Phi_h$. Different $\Phi_h$'s are also assumed independent. These assumptions are approximately valid for mesh topologies with random connection requests and spectrum assignment \cite{SC1,WC1}. 

Since no spectrum conversion capability is available, a package of $S$ contiguous spectrum slots that is free on all of the links of the shortest path connecting node $s$ to node $d$ should be assigned to the request. A spectrum slot is free on the shortest path with probability of $\prod_{h = 1} ^ {H} \Phi_h$. Let $Pr(S,F, \prod_{h = 1} ^ {H} \Phi_h)$ be the probability of finding at least $S$ contiguous spectrum slots on the shortest path. This looks like having at least $S$ consecutive head in $F$ coin flips where the probability of a head is $\prod_{h = 1} ^ {H} \Phi_h$.  If we define $\rho = \prod_{h = 1} ^ {H} \Phi_h$ then $Pr(S,F, \rho)$ can recursively be computed as follows \cite{SC1, Prob_Book}:
\begin{align}\label{Pr_S_F_Rho}
Pr(S,F,\rho) = \sum\limits_{j=1}^{S}Pr(S,F-j,\rho)\rho^{j-1}(1-\rho) + \rho^S
\end{align}
where initial terms are $Pr(S, F, \rho) = 0, \forall F < S$. Finally, the blocking probability equals to the complement of the connection request establishment so:
\begin{equation}\label{Non}
P_{B}^{sd}(S) = 1 - Pr(S,F,\prod_{h=1}^{H}\Phi_h)
\end{equation}

\subsubsection{$H$-hop Lightpath Connection Request Blocking Probability in Second Scenario}\label{sec_III_A_2}
In this scenario, the spectrum continuity constraint can totally be neglected so, a connection request is blocked when no package of $S$ contiguous spectrum slots exists on a link of the path. Therefore:
\begin{equation}\label{Full_Full}
P_{B}^{sd}(S) = 1 - \prod_{h=1}^{H}Pr(S,F,\Phi_h)
\end{equation}

\subsubsection{$H$-hop Lightpath Connection Request Blocking Probability in Third Scenario}\label{sec_III_A_3}
\begin{figure}[t!]
\center{\includegraphics[scale=0.75]{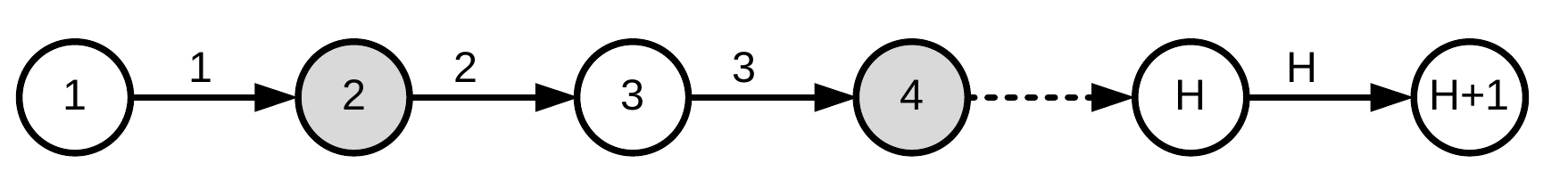}}
\center{\caption{\label{Numbering_SC} BVWXC and hop numbering format.}}
\end{figure}

Now, we assume some of the BVWXCs have Full architecture.  We use the notation $L = (L_1=1, L_2, L_3, ..., L_{\vert L\vert+1}, L_{\vert L\vert+2}=H+1)$ to show the distribution of SCBVWXCs over the shortest path, where $L_i,  i \in \lbrace 1, 2, ..., \vert L \vert+2 \rbrace$ shows $i$th element of vector $L$ and $\vert L\vert=N_{sc}$ is the number of SCBVWXCs. Note that the blocking probability of the lightpath is not related to the spectrum conversion capability in source and destination BVTs and consequently, $L_i, i \in \lbrace 2, 3, ..., \vert L \vert+1 \rbrace$ can take index values between $2$ and $H$. As example, $L = (1, 2, 4, H+1)$ for Fig. \ref{Numbering_SC} means that the lightpath begins at BVT indexed $1$, ends at BVT with index $H+1$ and $2$nd and $4$th BVWXCs have full spectrum conversion capability. 
A connection request is correctly established if $S$ contiguous spectrum slots are found on each sub-path between SCBVWXCs. Therefore, the blocking probability is:
\begin{equation}\label{Sparse_Full}
P_{B}^{sd}(S) = 1 - \prod_{k=1}^{N_{sc}+1}Pr(S,F,\prod_{h=L_k}^{L_{k+1}-1}\Phi_h)
\end{equation}

\subsubsection{$H$-hop Lightpath Connection Request Blocking Probability in Fourth Scenario}\label{sec_III_A_4}
In Share-per-Link structure, a bank of $N_{sc}$ SCBs is devoted to each output port and spectrum conversion capability is not always possible for all the lightpaths crossing a certain output port. Assume $N_{port_j}$ paths with total required spectrum slots $S_{port_j}$ cross the output port $j$ of a Share-per-Link SCBVWXCs. A typical path that crosses the output port $j$ averagely needs $\frac{S_{port_j}}{N_{port_j}}$ spectrum slots. This typical path doesn't need spectrum conversion with probability $\Phi_{port_j}^{\frac{S_{port_j}}{N_{port_j}}}$ so, an SCB will averagely be free for an incoming request if the number of paths requiring spectrum conversion is less than the number of embedded SCBs. Consequently,  at least one SCB is available with probability:
\begin{equation}\label{link_arch}
P^{port_j}_{node_i} = \sum_{k = 0}^{N_{sc}-1} {N_{port_j} \choose k} (1-\Phi_{port_j}^{\frac{S_{port_j}}{N_{port_j}}})^k \Phi_{port_j}^{\frac{S_{port_j}}{N_{port_j}} (N_{port_j}-k)}
\end{equation}
Again, we use the notation $L = (L_1=1,L_2, L_2, ..., L_{\vert L\vert+1}, L_{\vert L\vert+2}=H+1)$ to show how spectrum conversion capability is distributed among the shortest path. The power set of $L$ is defined as:
\begin{align}\label{power_set}
\nonumber & \mathcal{POW}(L) = \lbrace (1,H+1), (1,L_2,H+1), (1,L_3,H+1), \\
& ..., (1,L_2, L_3,H+1), (1,L_2, L_4,H+1), ..., L \rbrace
\end{align}
We define $\mathcal{U}_l(S)$ as the probability of successful establishment of the lightpath when SCBVWXCs are distributed according to $l$ and all of them participate in path establishment. $\mathcal{V}_l$ is the probability that SCBVWXCs distributed by $l$ are free to be used in the path establishment. Now, $\mathcal{U}_l(S)\mathcal{V}_l$ is the probability that all SCBVWXCs of the distribution $l$ successfully contribute to the path establishment. The path is successfully established with summing all of the possible $\mathcal{U}_l(S)\mathcal{V}_l$ of the given distribution $L$. Consequently:
\begin{equation}
P_B^{sd}(S) = 1 - \sum\limits_{l \in \mathcal{POW}(L)} \mathcal{U}_{l}(S) \mathcal{V}_{l}
\end{equation}
One can show that $\mathcal{U}_l(S)$ is recursively approximated as follows:
\begin{align}\label{link_recursive}
& \mathcal{U}_{(1,H+1)}(S) =  Pr(S,F,\prod_{h=1}^{H}\Phi_h) \\
\nonumber & \mathcal{U}_l(S) \approx  r\Big(\prod_{k=1}^{\vert l \vert-1}Pr(S,F,\prod_{h=l_k}^{l_{k+1}-1}\Phi_h)-\sum\limits_{l^\prime \in \mathcal{POW}(l) - l} \mathcal{U}_{l^\prime}(S)\Big)
\end{align}
where $r(\cdot)$ is the ramp function defined as:
\begin{equation}\label{ramp}
r(x)=
\begin{cases}
x &\text{$x \geqslant 0$}\\
0 &\text{$x < 0$}
\end{cases}
\end{equation}
And $\mathcal{V}_l$ is obtained by:
\begin{align}\label{sc_link_free}
\mathcal{V}_{(1,H+1)} =  1, \mathcal{V}_l =  \prod_{k=2}^{\vert l \vert-1}P^{port_{sd}}_{node_{l_k}}
\end{align}
where $port_{sd}$ is the output port of each SCBVWXC crossed by the connection request.

\subsubsection{$H$-hop Lightpath Connection Request Blocking Probability in Fifth Scenario}\label{sec_III_A_5}
This scenario is the same as the previously discussed situation except that the probability of having at least one SCB available is different. In Share-per-Node architecture the spectrum conversion bank is shared among all $N_{node_i}$ paths crossing the SCBVWXC. If $N^{port}_{node_i}$ is the number of the output ports, a crossing path averagely sees a spectrum slot free with the probability $\Phi_{node_i}$:
\begin{equation}
\Phi_{node_i} = \sum_{j = 1}^{N^{port}_{node_i}}\frac{N_{port_{j}}}{N_{node_i                                                                                     }}\Phi_{port_{j}}
\end{equation}
Defining $S_{node_i}$ as the total number of the spectrum slots required by $N_{node_i}$ crossing paths, the probability of having at least one SCB free is:
\begin{equation}
P_{node_i} = \sum_{k = 0}^{N_{sc}-1} {N_{node_i} \choose k} (1-\Phi_{node_i}^{\frac{S_{node_i}}{N_{node_i}}})^{k} \Phi_{node_i}^{\frac{S_{node_i}}{N_{node_i}}(N_{node_i}-k)}
\end{equation}
The remaining way is straight and the blocking probability is:
\begin{equation}
P_B^{sd}(S) = 1 - \sum\limits_{l \in \mathcal{POW}(L)} \mathcal{U}_{l}(S) \mathcal{V}_{l}
\end{equation}
where $\mathcal{U}_l(S)$ is recursively approximated by \eqref{link_recursive} and $\mathcal{V}_l$ is calculated as follows:
\begin{align}
\mathcal{V}_{(1,H+1)} =  1, \mathcal{V}_l =  \prod_{k=2}^{\vert l \vert-1} P_{node_{l_k}}
\end{align}

\subsubsection{$H$-hop Lightpath Connection Request Blocking Probability in Sixth Scenario}\label{sec_III_A_6}
Obviously, Empty and Full distributions are special cases of Sparse distribution. To provide a general description for the blocking probability, we consider Sparse distribution and allow each SCBVWXC to arbitrarily have one of the introduced architectures. Now, connection request blocking probability for distribution $L$ equals to:
\begin{equation}
P_B^{sd}(S) = 1 - \sum\limits_{l \in \mathcal{POW}(L)} \mathcal{U}_{l}(S) \mathcal{V}_{l}
\end{equation}
where $\mathcal{U}_l(S)$ is recursively approximated by \eqref{link_recursive} and $\mathcal{V}_l$ is:
\begin{align}\label{gen_form}
& \mathcal{V}_{(1,H+1)} =  1 ,\mathcal{V}_l =  \prod_{k=2}^{\vert l \vert-1} P_{sd_{l_k}}
\end{align}
$ P_{sd_{l_k}}$ in \eqref{gen_form} is related to SCBVWXC architecture and equals to:
\begin{equation}
P_{sd_{l_k}} = 
\begin{cases}
P_{node_{l_k}} &\mbox{for } \hspace{2 mm} \mathrm{Share-per-Node}\\
P^{port_{sd}}_{node_{l_k}} &\mbox{for }  \hspace{2 mm} \mathrm{Share-per-Link}\\
1 &\mbox{for }  \hspace{2 mm} \mathrm{Full}
\end{cases}
\end{equation}
\subsection{Average Network Blocking Probability in OFDM-based EONs}\label{sec_III_B}
Assume a network topology characterized by $G(V,E)$. The average connection request blocking probabilities, i.e. $\overline{P_{sd}}$'s and the average network blocking probability $P_B$ are defined as:
\begin{equation}\label{Pb_con}
\overline{P_{B}^{sd}} = \sum\limits_{S=1}^{\infty} P_{sd}(S)P_{B}^{sd}(S)
\end{equation}
\begin{equation}\label{Pb_net}
P_B = \frac{\sum\limits_{sd} R_{sd}T_{sd}\overline{P_{B}^{sd}}}{\sum\limits_{sd} R_{sd}T_{sd}}
\end{equation}
Expression \eqref{Pb_con} can be used to calculate the average connection request blocking probabilities if $\Phi_h$'s of the links are known. We estimate $\Phi_h$'s as follows:
\begin{equation}\label{Phi_h}
\Phi_h \approx 1-\min \lbrace \frac{\sum\limits_{sd: h \in sd} R_{sd}T_{sd}S_{sd}(1-\overline{P_{B}^{sd}})}{F}, 1 \rbrace
\end{equation}

Based on \eqref{Pb_con}, \eqref{Pb_net} and \eqref{Phi_h}, the iterative algorithm shown in Alg. \ref{alg1} can be used to provide an estimate of the average network blocking probability. At the first line of the algorithm, a temporary variable $P_{temp}$ with initial value of $-1$ is defined. Then, average connection request blocking probabilities $\overline{P_{B}^{sd}}$'s and average network blocking probability $P_B$ are initialized by random values chosen from the interval $[0,1]$. In the main loop of the algorithm, the current value of $P_B$ is backed up to $P_{temp}$ and then, $\Phi_h$'s are computed using \eqref{Phi_h}. Next, we calculate the values of $\overline{P_{B}^{sd}}$'s and $P_B$ according to the updated values of $\Phi_h$'s and equations \eqref{Pb_con} and \eqref{Pb_net}. The loop cycles until the difference between two successive values of $P_B$ becomes lower than a desired threshold $\epsilon$ (or the number of iterations reaches a predefined limit).
\begin{algorithm}[t!]
\caption{Calculate Average Network Blocking Probability}\label{alg1}
\begin{algorithmic}[1]
\Input{connection request specifications, a desired threshold $\epsilon$}
\Output{average network blocking probability $P_{B}$}
\Hline{\hspace{-.6 cm}\hrulefill}
\State $P_{temp}\longleftarrow -1$;
\State $P_{B} \longleftarrow rand[0,1]$;
\For{all connection requests $sd$}
\State $\overline{P_{B}^{sd}} \longleftarrow rand[0,1]$;
\EndFor
\While{$|P_{B}-P_{temp}| > \epsilon$}
\State  $P_{temp}\longleftarrow P_{B}$;
\For{all links $h$}
\State  update $\Phi_h$ using \eqref{Phi_h};
\EndFor 
\For{all connection requests $sd$}
\State  update $\overline{P_{B}^{sd}}$ using \eqref{Pb_con};
\EndFor
\State  update $P_{B}$ using \eqref{Pb_net};
\EndWhile
\end{algorithmic}
\end{algorithm}

\section{SCBVWXC Placement Algorithm}\label{sec_IV}
Assume $K$ BVWXCs of a given network topology can have one of the introduced architectures for SCBVWXCs to decrease the average network blocking probability. The question is that what is the best way of distributing SCBVWXCs to get the minimum average network blocking probability? A simple way is to test all ${\vert V \vert \choose K} K!$ possible cases and choose the best one in terms of the average network blocking probability but, it may last long or be computationally impossible \cite{WC_Placement}. Here, we take a heuristic approach to provide a sub-optimum but fast-achieved solution. Our heuristic algorithm is an extension of the heuristic procedure proposed in \cite{WC_Placement} and its steps are summarized in Alg. \ref{alg2}. To provide an estimate of the spectrum conversion capability of the $k$th architecture, we define a quantity named  $N^{k}_{eff}$. If the number of SCBs embedded in the $k$th architecture is $N_{sc}^k$, the mean value of the output ports of the BVWXCs in the network is $N_{port}$ and optical fibers contain $F$ spectrum slots, the value of $N^{k}_{eff}$ is given by:
\begin{equation} \label{sc_merit}
N^{k}_{eff} = 
\begin{cases}
N^k_{sc} &\mbox{if }  \hspace{2 mm} \mathrm{Full}\\
\frac{N_{sc}^k}{F} & \mbox{if }  \hspace{2 mm} \mathrm{Share-per-Link}\\
\frac{N_{sc}^k}{N_{port}F} & \mbox{if }  \hspace{2 mm} \mathrm{Share-per-Node}
\end{cases}
\end{equation}
In lines $1$ to $4$ of the heuristic algorithm, SCBVWXC architectures are decreasingly sorted according to the computed values of $N^{k}_{eff}$ and labeled from $1$ to $K$. Now assume that $G^{(k)}(V,E)$ shows the network topology at the beginning of $k$th iteration of the main loop in Alg. \ref{alg2}. In $k$th iteration, SCBVWXC $k$ is placed at each simple BVWXC (which has no spectrum conversion capability) of the network topology $G^{(k)}(V,E)$ and its corresponding average network blocking probability is computed using Alg. \ref{alg1}. At the end of iteration $k$, the SCBVWXC $k$ is placed at the node that corresponds to the minimum average network blocking probability among all the inspected locations and then, the network topology graph is updated to $G^{(k+1)}(V,E)$. Finally, the heuristic algorithm terminates after $K$ iterations.
\begin{algorithm}[t!]
\caption{SCBVWXC Placement Algorithm}\label{alg2}
\begin{algorithmic}[1]
\Input{network topology, connection request specifications, $K$ SCBVWXCs }
\Output{distribution of SCBVWXCs over the network topology}
\Hline{\hspace{-.6 cm}\hrulefill}
\For{all SCBVWXC architectures $k$}
\State compute $N^{k}_{eff}$ using \eqref{sc_merit};
\EndFor
\State sort SCBVWXC architectures decreasingly according to the values of $N^{k}_{eff}$ and number them from $1$ to $K$;
\State $G^{(0)}(V,E) \longleftarrow G^(V,E) $;
\For{all SCBVWXCs $k$}
\State $P_{temp} \longleftarrow 1$;
\State $ N_{temp} \longleftarrow 0$;
\For{all simple BVWXCs without spectrum conversion capability $n$}
\State place $k$th architecture in $n$th simple BVWXC of the network topology $G^{(k)}(V,E)$;
\State  compute $P_{B}$ using Algorithm \ref{alg1};
\If{$P_{B} < P_{temp}$}
\State $P_{temp} \longleftarrow P_{B}$;
\State $ N_{temp} \longleftarrow n$;
\EndIf
\EndFor
\State place $k$th SCBVWXC in node $N_{temp}$ of the network topology $G^{(k)}(V,E)$;
\State update network topology to $G^{(k+1)}(V,E)$;
\EndFor
\end{algorithmic}
\end{algorithm}

The computational complexity of the proposed heuristic algorithm is $\vert V \vert K - 0.5K(K-1)$ which is practically less than the brute force search complexity ${\vert V \vert \choose k} K!$. 
\section{Simulation Results}\label{sec_V}
\begin{figure}[t!]
\center{\includegraphics[scale=0.65]{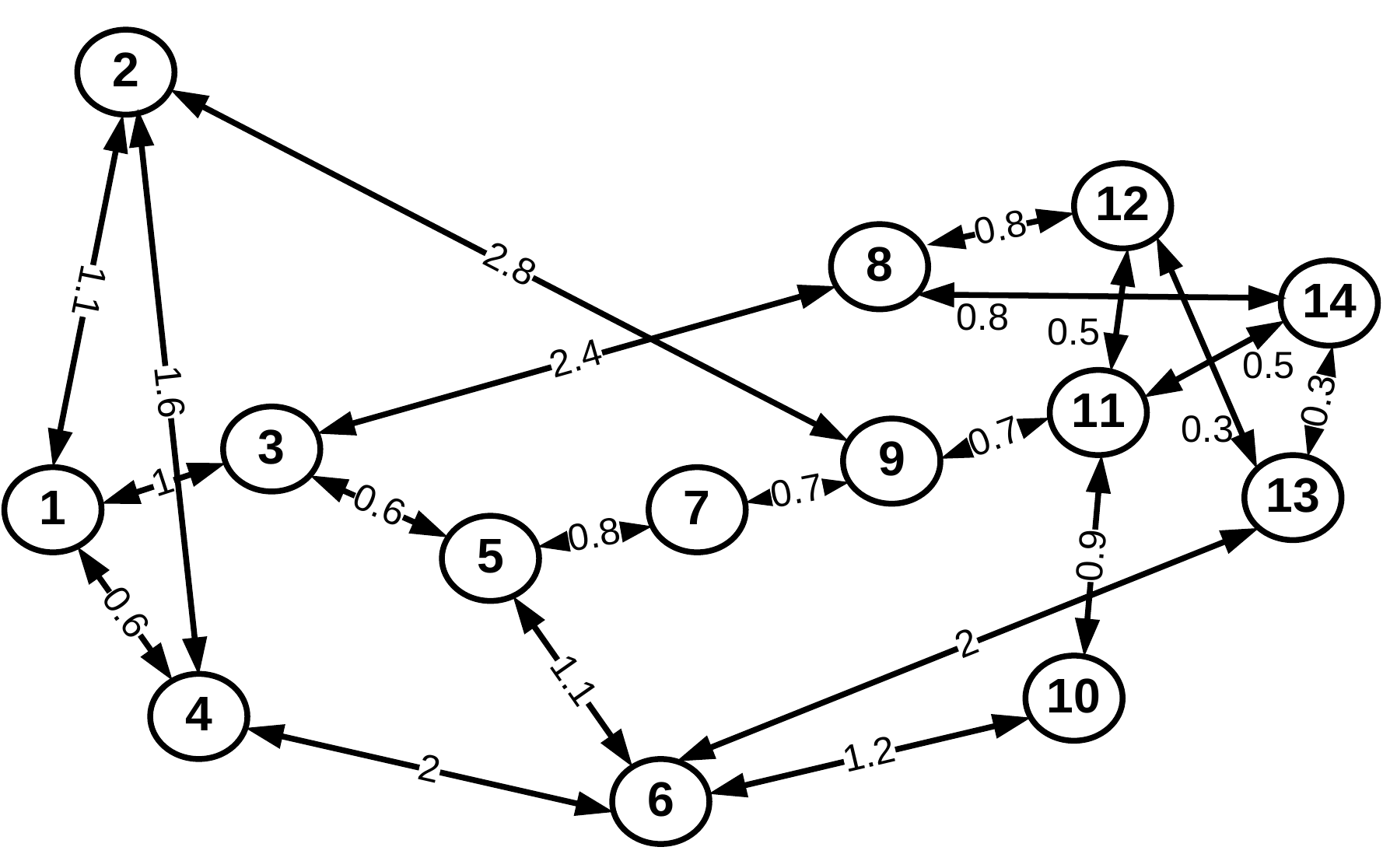}}
\center{\caption{\label{NSF} NSF network topology with $14$ nodes and $21$ bi-directional links. The number on each link indicates its corresponding weight \cite{WC_Placement}.}}
\end{figure}

\begin{figure}[t!]
\center{\includegraphics[scale=0.65]{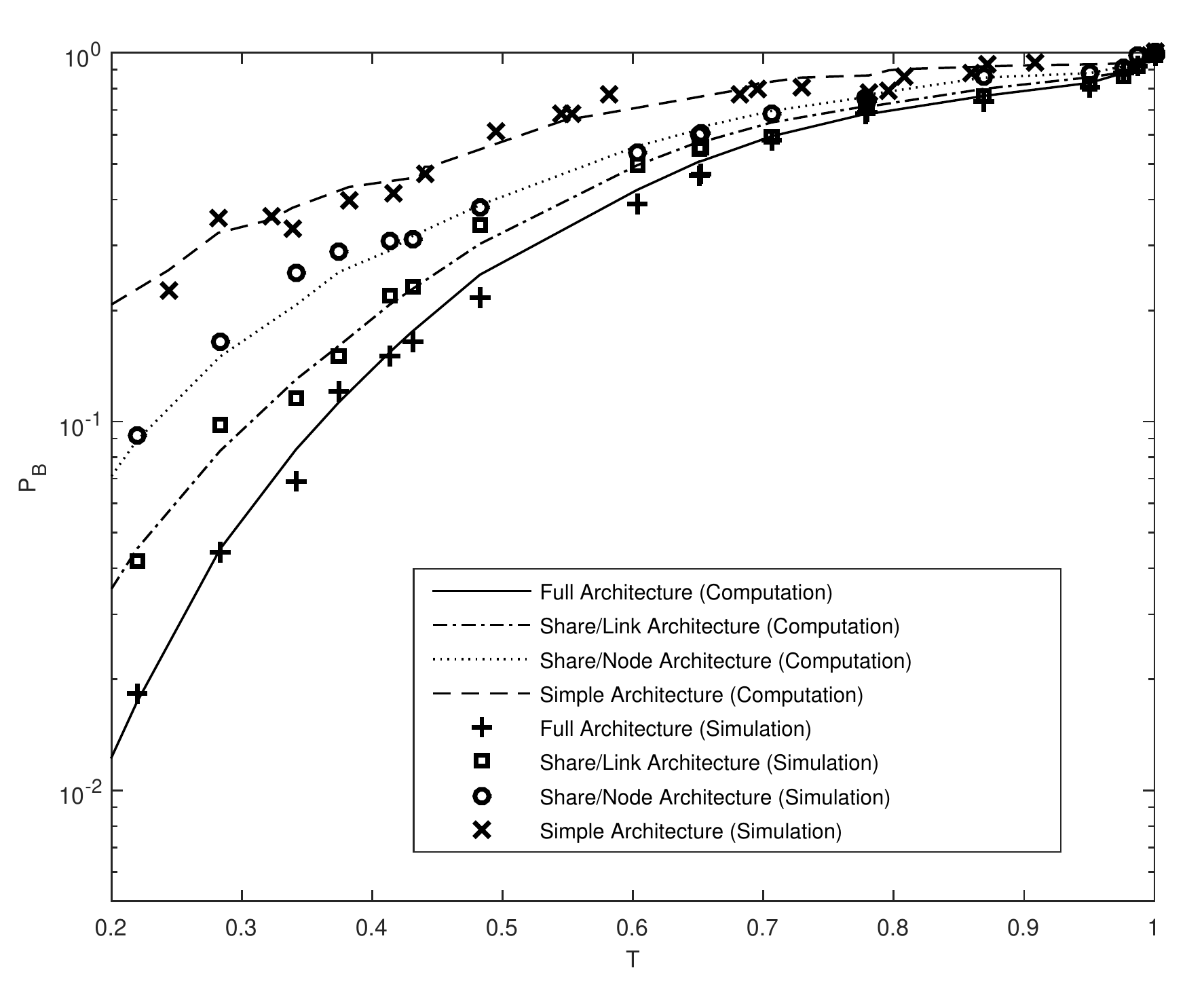}}
\center{\caption{\label{Sim1} Average network blocking probability $P_B$ of NSF network topology in terms of traffic $T$ for four different scenarios distinguished by various architectures of BVWXCs. "Full Architecture" means that all BVWXCs have Full architecture.  "Share/Link Architecture" and "Share/Node Architecture" mean that all BVWXCs have Share-per-Link and Share-per-Node architecture, respectively. By "Simple Architecture" we mean that there is no spectrum conversion capability in the network. The word "Computation" shows that the lines are plotted using the proposed computational framework while the word "Simulation" indicates that the markers are resulted from simulation. }}
\end{figure}

Consider NSF network topology \cite{WC_Placement} with $14$ nodes and $21$ bi-directional links shown in Fig. \ref{NSF}. 
We assume any pair of nodes in the network has a shortest path-routed connection request characterized by the mentioned notation $C_{sd} = (R_{sd}, T_{sd}, S_{sd})$ with a fixed number of required spectrum slots $S_{sd}$. The mean values of $R_{sd}$'s and $T_{sd}$'s and the fixed values of $S_{sd}$'s are uniformly chosen from pre-defined intervals such that a certain network traffic value with the following definition is obtained: 
\begin{equation}\label{traffic}
T = \frac{\sum\limits_{sd} R_{sd}T_{sd}S_{sd}N_{sd}^{hop}}{N_{link}^{net}F}
\end{equation}
where $N_{sd}^{hop}$ is the number of hops in the shortest path connecting source $s$ and destination $d$ and $N_{link}^{net}$ is the number of directional links in the network topology. Considering NSF network, Fig. \ref{Sim1} shows the average network blocking probability $P_B$ in terms of traffic $T$ for four different scenarios distinguished by various architectures of BVWXCs. The lines are plotted using the proposed computational framework for calculating the average network blocking probability while the markers are resulted from simulating the network with desired configurations. Obviously, there is an acceptable match between the results of the proposed computational framework and simulation. The average network blocking probability is an ascending function of the traffic. For a fixed value of the traffic, the average network blocking probability can be improved by embedding spectrum conversion capability in BVWXCs. The most improvement is for Full architecture and the performance respectively decreases for Share-per-Link and Share-per-Node architectures.

\begin{figure}[t!]
\center{\includegraphics[scale=0.65]{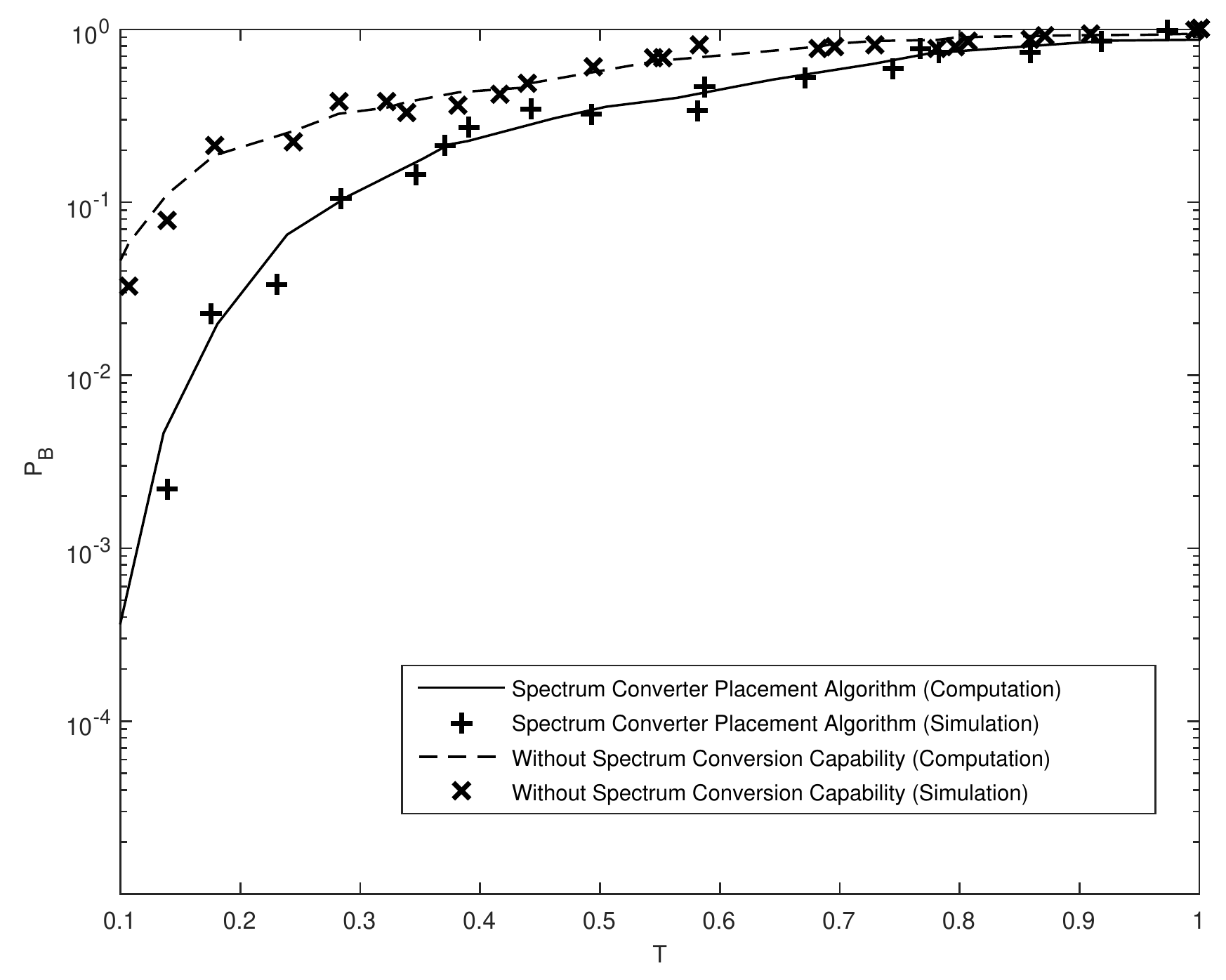}}
\center{\caption{\label{Sim2} Average network blocking probability $P_B$ of NSF network topology in terms of traffic $T$ before and after equipping the network with spectrum conversion capability based on the proposed SCBVWXC placement algorithm. The word "Computation" shows that the lines are plotted using the proposed computational framework while the word "Simulation" indicates that the markers are resulted from simulation. }}
\end{figure}

To evaluate the performance of the SCBVWXC placement algorithm, assume that we can equip three of the BVWXCs of the NSF network with spectrum conversion capability, two of them in Full architecture and one of them in Share-per-Node architecture with one shared SCB. The SCBVWXC algorithm places two Full architectures in $11$th and $6$th nodes while node $14$ is offered for placing the Share-per-Node SCBVWXC. Brute force search for the best distribution of these three SCBVWXCs results in the same solution which indicates that the heuristic SCBVWXC placement algorithm achieves the optimum solution in this case. Intuitively, the SCBVWXC placement algorithm nominates the crowded nodes for placing SCBVWXCs. Fig. \ref{Sim2} compares the average network blocking probability $P_B$ in terms of traffic $T$ before and after equipping the network with spectrum conversion capability based on the proposed SCBVWXC placement algorithm.
\section{Conclusion}\label{sec_VI}
A certain amount of spectrum conversion capability can be embedded in a given Elastic Optical Network (EON) topology to improve the network performance by decreasing connection request blocking probability. According to the way of spectrum conversion sharing in an optical Bandwidth-Variable Wavelength Cross-Connect (BVWXC), we introduce three architectures named Full, Share-per-Link and Share-per-Node for Spectrum-Convertible Bandwidth-Variable Wavelength Cross-Connects (SCBVWXCs). We consider Full, Sparse and Empty distributions of SCBVWXCs in an EON and propose a computational framework for calculating average network blocking probability covering various architectures and distribution methods of SCBVWCs. As another contribution, we propose a heuristic algorithm for distributing a limited number of SCBVWXCs in a given network topology such that the average network blocking probabilisty is minimized. Finally, we use simulation results to evaluate the performance of the mathematical and algorithmic achievements.

\end{document}